# Diffractive imaging of dissociation and ground state dynamics in a complex molecule

Kyle Wilkin[1,*], Robert Parrish[2,3,4,*], Jie Yang[2,3,*], Thomas J. A. Wolf[3], J. Pedro F. Nunes[1,5], Markus Guehr[3,6], Renkai Li[2], Xiaozhe Shen[2], Qiang Zheng[2], Xijie Wang[2], Todd J. Martinez[3,4], Martin Centurion[1]

[1]Department of Physics and Astronomy, University of Nebraska – Lincoln, 855 N 16th St., Lincoln, NE, 68588, United States.

[2]SLAC National Laboratory, 2575 Sand Hill Rd, Menlo Park, CA, 94025, United States.

[3]Stanford PULSE Institute, Menlo Park, CA, 94025, United States.

[4]Department of Chemistry, Stanford University, 333 Campus Dr., Menlo Park, CA, 94305, United States.

[5]Department of Chemistry, University of York, Heslington, York, YO10 5DD, UK.

[6]Institut für Physik und Astronomie, Universität Potsdam, Potsdam, 14476, Germany.

*These authors contributed equally to the work.

**Abstract** We have investigated the structural dynamics in photoexcited 1,2-diiodotetrafluoroethane molecules ($C_2F_4I_2$) in the gas phase experimentally using ultrafast electron diffraction and theoretically using FOMO-CASCI excited state dynamics simulations. The molecules are excited by an ultra-violet femtosecond laser pulse to a state characterized by a transition from the iodine $5p\perp$ orbital to a mixed $5p||$ σ hole and $CF_2^\bullet$ antibonding orbital, which results in the cleavage of one of the carbon-iodine bonds. We have observed, with sub-Angstrom resolution, the motion of the nuclear wavepacket of the dissociating iodine atom followed by coherent vibrations in the electronic ground state of the $C_2F_4I$ radical. The radical reaches a stable classical (non-bridged) structure in less than 200 fs.

## I. INTRODUCTION
### A. Background

The conversion of light into chemical and mechanical energy at the level of single molecules is an essential mechanism that has applications across multiple fields. In order to fully understand and accurately model these processes, it is crucial to observe them with atomic resolution on their natural timescale of femtoseconds. The photodissociation of 1,2-diiodotetrafluoroethane ($C_2F_4I_2$) is a nonconcerted reaction in which an iodine atom is eliminated quickly, within 200 fs, while the second iodine is eliminated on a longer timescale of tens of picoseconds [1-7]. The dynamics of this reaction have been studied extensively, both experimentally and theoretically, in order to determine whether the short-lived radical takes on a bridged structure where the iodine atom is equidistant between the two carbon atoms, or whether it takes a classical structure that resembles the parent molecule before one iodine atom is eliminated [8-13]. In picosecond X-ray diffraction experiments in solution, the photo-dissociation of $C_2F_4I_2$ was determined to lead to the classical radical structure, while experiments with $C_2H_4I_2$ showed that the radical took the bridged structure [10,14-16]. The classical structure for the intermediate $C_2F_4I$ in the gas phase was observed with Ultrafast Electron Diffraction (UED) with a resolution of a few picoseconds [1,3,7]. It is still an open question whether the bridged structure of $C_2F_4I$ in the gas phase exists on a shorter time scale. In addition, the previous experiments could not directly observe the formation of the radical and the ensuing dynamics because the combined spatiotemporal resolution was not sufficient to resolve the underlying nuclear motion.

Previous gas phase picosecond UED experiments have been used to capture the structure of short lived intermediate molecular states [1,3,17] and to determine 3D molecular structure from aligned molecules [18]. Although UED experiments in condensed matter samples were able to reach femtosecond resolution [19], in gas phase experiments the resolution had been limited to the picosecond timescale, mainly by the velocity mismatch between laser and electron pulses. Recently, it has become possible to achieve femtosecond resolution in gas phase UED using relativistic (MeV) electron pulses [20], with first experiments that observed rotational and vibrational nuclear wavepackets in diatomic molecules [21,22] and recent experiments that captured the dynamics in more complex reactions such as the passage of a nuclear wavepacket through a conical intersection and ring-opening dynamics [23,24]. Here we use UED to investigate the prompt photo-dissociation of $C_2F_4I_2$, the formation of the $C_2F_4I$ radical and the subsequent coherent dynamics in the electronic ground state. We have experimentally observed the nuclear wavepacket of the dissociating iodine atom, determined that the classical structure is formed within one vibrational period of the C-I bond (~200 fs), and measured coherent motion in the radical corresponding to vibrations and rotations of the isolated $CF_2$ group. The observed structure of the radical and coherent motion are in agreement with FOMO-CASCI [25-27] excited state dynamics simulations with 12 electrons in 8 orbitals.

### B. Static Gas Electron Diffraction

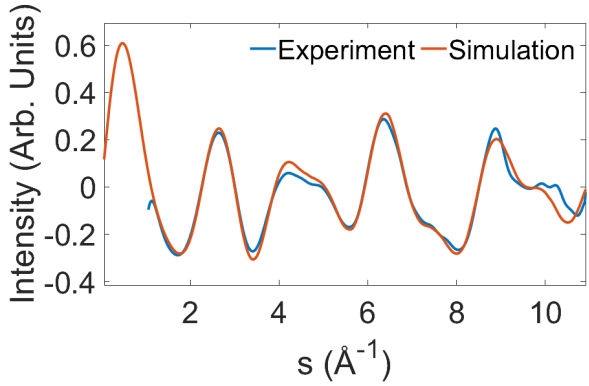

**FIG. 1.** Weighted Legendre projection of the modified molecular scattering pattern (*sM*) for both experiment and simulations.

The total scattering from a randomly oriented distribution of molecules in the gas phase can be written as:

$$I_{total} = \sum_i^N |f_i(s)|^2 +$$

$$\sum_i^N \sum_{j \neq i}^N f_i(s) f_j(s) \cos(\eta_i - \eta_j) \frac{\sin(r_{ij}s)}{r_{ij}s} \quad (1)$$

where $s = \frac{4\pi}{\lambda}\sin\left(\frac{\theta}{2}\right)$ is the momentum transfer of the scattered electrons, $\lambda$ is the de Broglie wavelength of the electrons, $\theta$ is the scattering angle, $f_i$ and $\eta_i$ are the scattering amplitude and phase for the *i*-th atom respectively, $r_{ij}$ is the distance between the *i*-th and *j*-th atom, and $N$ is the total number of atoms in the molecule. The first term on the right-hand side of equation 1 is the atomic scattering intensity, $I_{atomic}$, and contains no structural information. The second term, the molecular scattering $I_{molecular}$, is a sum over the interference pattern created by each atom pair in the molecule. The modified molecular scattering intensity $sM(s) = s \times \frac{I_{molecular}}{I_{atomic}}$ is used to normalize the diffracted intensity which decreases very rapidly with scattering angle. In the case of photo-excitation, the molecules are not randomly oriented due to the angular selectivity of the process. In this case, $sM$ can still be retrieved by first projecting the diffraction pattern onto the Legendre polynomials weighted with $|\sin(\varphi)|$ and keeping only the zeroth order [28].

The weighted Legendre projection used to approximate $sM_n(s)$ is given by

$$sM_n(s) \approx \int P_n(\cos(\varphi))\, sM(s,\varphi)\, |\sin(\varphi)|\, d\varphi \quad (1)$$

Where $P_n$ is the n-th order Legendre polynomial, $\varphi$ is the azimuthal angle in the diffraction image, and $sM(s,\varphi)$ is the experimental modified scattering pattern as described above. The isotropic contribution of the $sM$ corresponds to $sM_0$ [29]. The $sM_0$ projection is used when referencing $sM$ in this manuscript. Further discussion of the Legendre projection can be found in Appendix a.

Figure 1 shows the static experimental and theoretical $sM(s)$. The diffraction patterns are pre-processed to remove noise from dark current, scattered laser light, and random non-scattering events. The details of the data analysis are described in Appendix b. The atomic scattering and background are removed from the data using standard methods from gas electron diffraction [1]. Experimental data for s < 1.2 Å$^{-1}$ is not available due to a hole in the detection apparatus to allow the non-scattered electrons through. The experiment and theory show quantitative agreement up to $s = 11$ Å$^{-1}$.

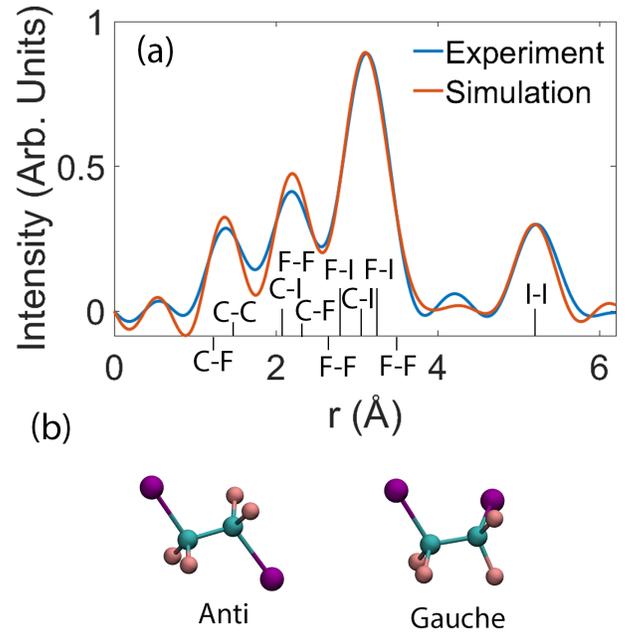

**FIG. 2.** (a) Static experimental and theoretical pair distribution function (*fr*) of the parent molecule $C_2F_4I_2$. All atomic distances in the anti conformer are labeled. (b) Structural models of the parent molecule $C_2F_4I_2$ for both the anti and gauche conformers.

The projection is then transferred into real space by a sine transform of the $sM$ giving the pair distribution function, *fr*. The *fr* contains information about the interatomic distances within the molecule and is defined by,

$$fr(r) = \int_0^{s_{max}} sM(s) \sin(sr)\, e^{-ds^2}\, ds$$

Where $r$ is distance in real space, $s_{max}$ is the largest maximum value of momentum transfer detected by the experiment, $d$ is a damping factor used to minimize the effect of the discontinuity at $s_{max}$. The ideal transform is an integral over all scattering angles so standard diffraction analysis techniques (see Appendix b for details) are used to compensate for any missing experimental scattering angles. Figure 2a shows the experimental and simulated $fr$. The peaks indicating atomic distances found in the parent molecule have a strong overlap in both position and amplitude.

$C_2F_4I_2$ has two stable structures in the ground state, anti and gauche. The conformers are characterized by their ICCI dihedral angle. The anti has $C_{2h}$ symmetry with a dihedral angle of 180° while the gauche has $C_2$ symmetry with a dihedral angle of 70°. Static models of the two are shown in Fig. 2b. The ratio of the two conformers is dictated by the temperature of the sample with the anti conformer the more stable structure [30]. The experimental static $sM$ was used to determine the conformer percentage in the sample. Using the least squares method the conformer ratio was determined to be 9:1 anti to gauche. Using the van 't Hoff parameters from [30] we determine the temperature of the sample to be 220 Kelvin. This temperature is within the expected range for an expanding beam of $C_2F_4I_2$ seeded with one Bar of helium.

To elucidate the changes in the diffraction pattern that happen after the excitation by the laser we employed the diffraction-difference method [31], where $\Delta sM(s,t) = sM(s,t) - sM(s,-\infty)$ . Similarly, $\Delta fr(r,t) = fr(r,t) - fr(r,-\infty)$.

## II. EXPERIMENTAL SETUP

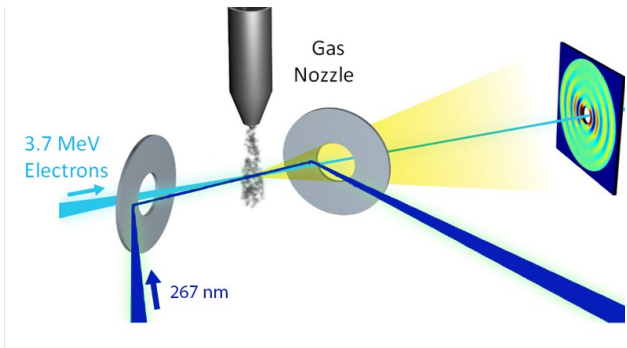

**FIG. 3.** (a) Illustration of the interaction region of the experimental setup.

The UED experiment reported here was carried out using the MeV UED setup at SLAC, sketched in Fig. 3 and described elsewhere in detail [23]. Briefly, the electron and laser beam are approximately collinear as they traverse a gas jet containing $C_2F_4I_2$ molecules. The electrons are accelerated to 3.7 MeV in an RF gun at a repetition rate of 120 Hz, with each pulse containing 20k-25k electrons on target. The pump laser is linearly polarized with a central wavelength of 264.5 nm and FWHM bandwidth of 1.3 nm. The laser pulse energy is 2.52 μJ focused to a 308 μm by 378 μm spot size giving a fluence of 2.2 $\frac{mJ}{cm^2}$ at the interaction region. The time duration of the laser pulse is expected to be 80 fs FWHM. The laser propagates through the sample at a 5° angle with respect to the electrons ensuring the velocity of the laser and electrons through the sample are approximately equal. The excitation percentage was determined to be 6%, based on the magnitude of the changes in the diffraction signal.

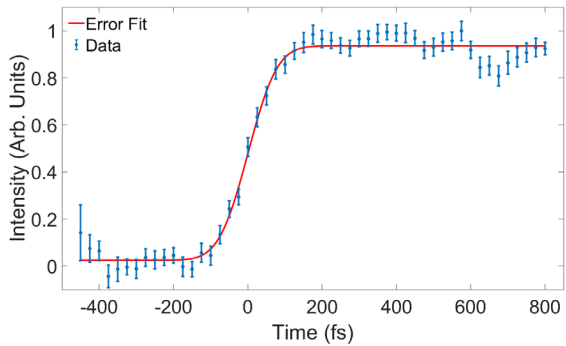

**FIG. 4.** Changes in the $\Delta sM$ fit with an error function to determine the instrument response time.

The measured instrument response time is (145 fs) is a convolution of the temporal resolution of the setup with the response of the molecular system to the laser excitation. Thus it can be considered as an upper limit of the temporal resolution of the experiment The instrument response time was determined from the largest changes in the $\Delta sM$ in the region 1.5 Å$^{-1}$ < s < 3 Å$^{-1}$ (Fig. 4). The signal was fit to an error function and the width of the associated Gaussian gives the instrument response.

The scattered electrons are collected by a phosphor screen which is imaged onto a CCD camera. The phosphor screen has a hole in the center to transmit electrons that are not scattered by the sample. Images were recorded over 2 days with several time scans per day. Each time scan recorded approximately 64 images at different relative time delays between the laser and electron pulses. An approximate value for $t$=0, where the laser and electrons arrive at the sample simultaneously, was found by increasing the intensity of the pump laser to create a plasma which distorts the transmitted electron beam. Each scan includes images at relative delays between -10 ps to 30 ps with steps of 25 fs taken close to $t=0$ and larger steps outside of ±1 ps. Most of the data covers the range between -300 fs and 800 fs and is the focus of this manuscript. In total each time step has approximately 1000 seconds of accumulated data.

The $C_2F_4I_2$ sample was purchased from SynQuest Laboratories at 97% purity. The sample was heated in a bubbler and delivered to the chamber through a gas line flowing helium at a pressure of 1 bar. $C_2F_4I_2$ is injected into the sample chamber via a 100 μm pulsed nozzle. The bubbler, gas line, and nozzle are heated to 60 °C, 100 °C, and 120 °C respectively. The repetition rate of the nozzle is set to match the repetition rate of the electrons at 120 Hz and had an opening time of 200 μs. The gas jet is approximately 300 μm FWHM at the interaction region.

## III. THEORETICAL METHODS

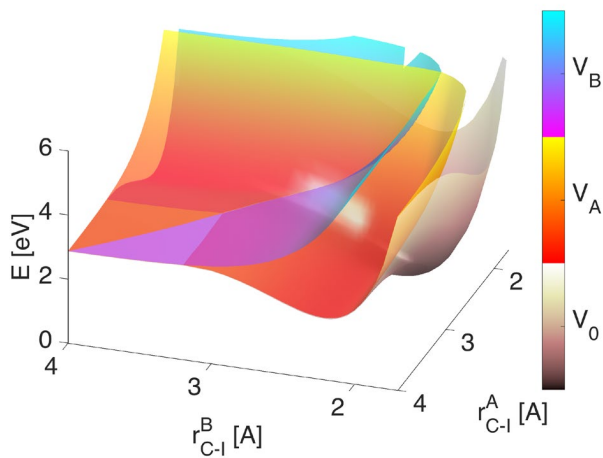

**FIG. 5.** Potential energy surfaces corresponding to the rigid stretching of the two C-I bonds in $C_2F_4I_2$.

### A. FC Point:

The anti and gauche conformers of C2F4I2 were optimized on $S_0$ at the FC point using FOMO-CASCI(12e,8o,kT = 0.2   au)-D3/LANL2DZ with Gaussian smearing, using TERACHEM and DLFIND in delocalized internal coordinates. The LANL2DZ basis set uses effective core potentials (ECPs) for the $1s^22s^22p^63s^23p^63d^{10}4s^24p^64d^{10}$ electrons of I, and then uses a DZ-quality basis set for the $5s^25p^5$ electrons - all other electrons on C and F atoms are treated explicitly. Overall, 92 electrons are treated with ECPs, 62 are treated explicitly, and there are 70 total atomic orbitals (AOs). The Hessian was computed at the same level of theory via finite difference of analytical gradients. All harmonic vibrational frequencies are real, with lowest vibrational frequencies of 54 (45) cm$^{-1}$ corresponding to twisting of the I-C-C-I dihedral and highest vibrational frequencies of 1412 (1143) cm$^{-1}$ corresponding to C-C bond stretch for anti (gauche) conformers. Note the remarkably small maximum vibrational frequencies of <1500 cm$^{-1}$ in this system, compared to >3100 cm$^{-1}$ typically encountered in hydrogen-containing systems - this allows for the safe application of a large timestep for dynamics simulations of 1 fs in $C_2F_4I_2$ compared to the 0.5 fs timestep usually used in hydrogen containing systems.

### B. Initial Conditions

For each conformer, ground-state initial conditions for forthcoming trajectory-based excited-state dynamics simulations were drawn from a harmonic Wigner distribution at 0 K, using the FOMO-CASCI level of theory described above. Note that the nature of the experimental sample preparation with gas jet expansion precludes the straightforward assignment of an equilibrium temperature to the experimental initial conditions.

### C. Potential Energy Landscape

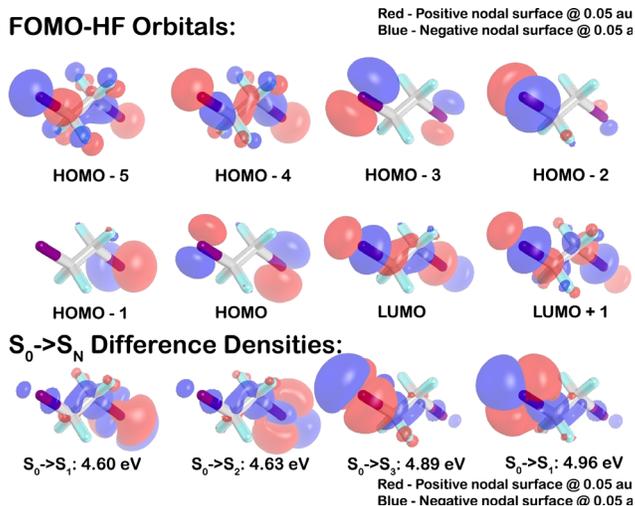

**FIG. 6.** Example FOMO-CASCI(12e,8o,kT=0.2)-D3/LANL2DZ active orbitals (top two rows) and S0->SN difference densities (bottom row) for a randomly selected anti-conformer T=0K Wigner sample (#0000).

To describe the excited state dynamics of $C_2F_4I_2$, we must locate a level of electronic structure theory that is capable of accurately describing the excited state landscape in the vicinity of the FC region, the known photodissociation products of $C_2F_4I$ + I and $C_2F_4$ + 2I, and the intermediate regions between these two limits. The proper treatment of two asymptotically separated I radicals requires all possible configurations of 5 electrons in the 5p orbitals of both dissociated I fragments, i.e., requiring a CASCI(10e,6o) active space. The treatment of the relevant excited states at the FC point requires excitations from the distorted 2 × $5p\perp$ orbitals in the C-I bond to the sigma-hole $5p\parallel$ + CF$_2$• radical antibonding orbital on both C-I bonds, i.e., requiring a CASCI(8e,6o) active space (2 particle orbitals). The union between the required numbers of holes and particles in these active spaces is CASCI(10e,7o). However, we have found that CASCI(10e,7o) has an improper symmetry-broken solution at medium-to-large separations of both I atoms - the π and π* orbitals of the C$_2$F$_4$ moiety are retained in the active space for large regions of the doubly dissociating potential energy surface, precluding the selection of the proper (10e,6o) active space on the I atoms (i.e., one too few I 5p atoms is included in the active space). This manifests as an artificial barrier to double dissociation and an incorrect 6-fold degeneracy of the doubly-dissociated system. Increasing the active space to (12e,8o) ameliorates this issue entirely, and does not significantly affect the potential surface near the FC point. Thus we have selected FOMO-CASCI(12e, 8o, kT = 0.2 au) -D3/LANL2DZ for the preferred level of theory in this study. See Fig. 6 for an example of the active orbitals and ground-to-excited difference densities at a representative anti-conformer Wigner sample.

An informative view of the potential energy landscape is obtained by varying both the C-I bond distances (labeled A and B) along the bond axis from the FC-optimized geometry (the rest of the molecule is frozen during this study). This is depicted for the anti conformer in Fig. 5, using our selected FOMO-CASCI(12e,8o,kT = 0.2 au)-D3/LANL2DZ level of theory. Note that the adiabatic version of this potential energy landscape is symmetric, by definition, about the $r_C - I^A = r_C - I^B$ hyperplane. The ground state potential consists of a smooth, Morse-like well at the FC point with an optimum C-I bond length of 2.21 Å. The C$_2$F$_4$I + I dissociation channel is visible upon stretching either C-I bond.

Considering the S$_0$ asymptotic dissociation channels in isolation is misleading: inspection of the full singlet potential energy manifold reveals that the single dissociation channel is asymptotically 3-fold degenerate (corresponding to the 3 possible orientations p$_x$, p$_y$ or p$_z$ of placing the hole in the dissociated I fragment), and that the double dissociation channel is 9-fold degenerate (3 possible hole orientations for 2 × I fragments). If one were to consider triplet and/or quintet asymptotic states, the degeneracy would be even higher. Focusing on the singlet subspace, the way in which the higher-lying surfaces coalesce to the asymptotic dissociation channels is extremely interesting. Four key states are accessible from the FC point with the 4.66 eV pump used in the experiment. These are most easily discussed in a diabatic picture. For each C-I bond A, there are two essentially degenerate diabats $V_A^1$ and $V_A^2$. Near the FC point, these correspond to an excitation from either of the two occupied and distorted perpendicular 5p hole orbitals (those promoting the C-I bond) to a particle orbital which is a mixture of parallel σ-hole-type 5s/5p orbital and CF$_2$ radical orbital. At the FC point, all four of these states are within 0.10 eV of each other with excitation energies of 4.51 eV to 4.61 eV. The two states for each C-I bond differ in the orientation of the $5p\perp$ hole orbital.

Near the FC point, each of these excitations destabilizes the involved C-I bond, and applies a significant repulsive force along the C-I bond axis. This promotes dissociation to a CF$_2$+I moiety. The CF$_2^+$+I$^-$ dissociation channel is visible on the potential surface by stretching the non-excited C-I bond. While this is accessible from the FC point, this process is not favored, as there is negligible driving force toward this channel (significant force promoting this channel is only realized if the non-excited C-I bond is significantly compressed at the FC point). Therefore, the primary effect of photoexcitation is expected to be prompt dissociation to C$_2$F$_4$I + I. Note that the diabats for the A and B dissociations cross by definition at $r_C - I^A = r_C - I^B$ due to symmetry. Considering the adiabatic surfaces, these are sharply avoided crossings, which would correspond to rapid hole and particle transfer from one C-I bond to the other if one were to remain on the same adiabatic surface through such a crossing. Thus, it is expected that the dynamics will exhibit diabatic passage if started on the higher diabat (corresponding to exciting and then dissociating the more-compressed of the C-I bonds).

As one approaches the double-dissociation regime near $r_C - I^A > 3$ Å and $r_C - I^B > 3$ Å, a number of additional higher-lying states come down to intersect with S$_0$ to S$_5$, and produce the 9-fold degenerate coalesced ground state for C$_2$F$_4$ + 2I channel.

### D. Absorption Spectrum Computations

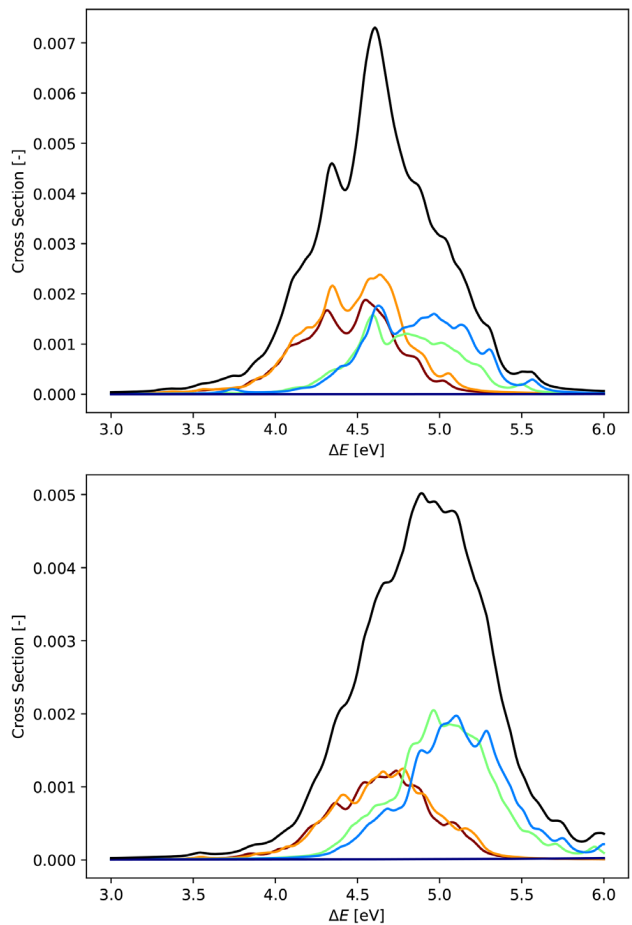

**FIG. 7.** Simulated absorption spectrum computed using FOMO-CASCI(12e,8o,kT = 0.2 au)-D3/LANL2DZ from 100 samples of 0 K harmonic Wigner distribution computed at the same level of theory. Each $S_0 \to S_N$ excitation is weighted by the transition-dipole-based oscillator strength and broadened by a Lorentzian with a width parameter of $\delta = 0.1$ eV. Top panel: anti. Bottom panel: gauche. Colors: dark red - $S_0 \to S_1$, orange - $S_0 \to S_2$, green - $S_0 \to S_3$, light blue - $S_0 \to S_4$, dark blue - $S_0 \to S_5$ (very bright transition, centered at ~ 7 eV), Black - total.

The absorption spectrum of $C_2F_4I_2$ (shown in Fig. 7) is calculated by computing vertical excitation energies at each Wigner sample, weighting by oscillator strength, broadening with a Lorentzian (width parameter $\delta \equiv 0.1$ eV), and summing the contributions from all excited states and Wigner samples. The anti and gauche spectra are similar. For the gauche states, the two diabats $V_A^1$ and $V_A^2$ or $V_B^1$ and $V_B^2$ differ from each other by an average of ~0.05 eV, while the two sets of diabats $V_A^{1/2}$ and $V_B^{1/2}$ differ from each other by an average of ~0.4 eV. For the anti Wigner samples, the average excitation energies are 4.35 (4.71) eV for the lower (upper) diabat pairs. For the gauche Wigner samples, the average excitation energies are 4.60 (5.02) eV for the lower (upper) diabat pairs. Also noteworthy is that there is rather small diversity in the oscillator strength weights. The relative standard deviation $\sigma_{Osc}/\mu_{Osc}$ is 0.55 (0.40) for all 400 oscillator strengths considered here for anti(gauche).

### E. Adiabatic Dynamics Simulations

Adiabatic dynamics simulations were conducted from the Wigner distributions above. In all cases, the dynamics were initiated on S1 and propagated for 1 ps using a timestep of 1 fs. Our main dataset consists of 100x trajectories initiated from a 0K Wigner distribution for anti and gauche conformers, and using electronic temperatures of 0.1, 0.2, and 0.3 au in FOMO-CASCI(12e,8o)-D3/LANL2DZ (600 trajectories total). 8 of 600 trajectories experienced SCF or CPSCF convergence failure before the end of the 1 ps time window, a 1.3% failure rate. This failure rate is likely to be insignificant relative to other uncertainties, so we ignore the missing frames from these trajectories in subsequent analysis.

### F. MS-CASPT Benchmarking

To probe the sensitivity of the results to differential dynamical electron correlation, limited XMS-CASPT benchmarks were performed at some of the rigid 2D scan geometries of the anti conformer based on SA-5-CASSCF(12e,8o)/LANL2DZ with XMS1 and a level shift of 0.3 au. 6 occupied electrons are frozen in the MS-CASPT2 computations to provide tractability in 32-bit Molpro. These agree semi-quantitatively with the FOMO-CASCI-D3(12e,8o,kT=0.2)/LANLDZ surfaces (agreements to within at worst a few tenths of 1 eV), but disagree significantly with the FOMO-CASCI-D3(10e,7o,kT=0.2)/LANL2DZ surfaces. Overall these results strongly indicate that differential dynamical electron correlation effects are insignificant for this system (at least within the limited LANL2DZ basis set) and validate the choice of electronic temperature of kT=0.2 au in the FOMO-CASCI computations.

### G. Photoexcitation and Dissociation

Photoexcitations driven by the 265 nm pump wavelength used in the experiment involve four relevant excited states ($S_1$-$S_4$) ranging from 4.51 eV to 4.61 eV at the FC point. These can be divided into two quasi-degenerate subsets corresponding to exciting each C-I bond from an occupied $5p\perp$ orbital to a mixed $5p\|$ $\sigma$ hole and $CF_2^\bullet$ antibonding orbital. The quasi-degeneracy of each pair of states stems from the 2x $5p\perp$ orbitals on each bond. Photoexcitation to one of these states results in a significant dissociative force being applied along the direction of the involved C-I bond, promoting prompt dissociation to $C_2F_4I$ + I. As this dissociation occurs, the

2-fold degenerate excited state coalesces with the ground state dissociation limit to form a 3-fold degenerate singly-dissociated $C_2F_4I$ + I (in the absence of spin-orbit coupling). Double dissociation to $C_2F_4$ + 2I is accessible from the excited FC point and involves higher 9-fold degeneracy. In an adiabatic picture the four excited states at the FC point appear to be weakly avoided crossings, e.g., as evidenced by sharp crossings in the adiabatic picture and by the localizability of the excited states to the two C-I bonds. The theoretical analysis shows a prompt dissociation of the second C-I bond after the first one in some of the trajectories, which is at odds with previous measurements of the reaction, where the second bond breaking takes place on a timescale of tens of picoseconds. This is a limitation of the simulations which we believe is caused by the absence of spin orbit coupling in the theory and the small energy difference between $C_2F_4I$ and $C_2F_4$. Thus, for comparison with experiments, we keep only the trajectories where a single C-I bond is broken in the first 800 fs and focus on the dynamics and structure of the $C_2F_4I$ radical.

### H. Discussion of Known Unknowns

There are a number of potential sources of error in the electronic structure and dynamics simulations. Here we discuss known sources of such errors and their potential likelihood/effects on the observable data, in decreasing order of perceived importance.

*Triplets/Intersystem Crossings/Spin-Orbit Coupling*:

The leading source of error in the above model is likely to be lack of inclusion of spin-orbit coupling. The presence of I atoms in the system is likely to significantly alter the potential surface of the dissociated species, e.g., by the ~0.93 eV splitting between isolated I and I* species. Previous experimental results by Zewail suggest that the majority (70%) of the dissociation products are I* species [2]. The model error due to lack of inclusion of this effect is likely to be excessive apparent available kinetic energy during the early stages of the dynamics (as dissociation to I* would remove 0.93 eV of the excess energy from the pump photon). This would manifest as an increased branching ratio for concerted double dissociation, which is what we observe. Therefore, we separate out the part of the wavepacket pertaining to single dissociation events to compute detailed observables for this channel, without attempting accurate determination of the branching ratio (this would require full inclusion of spin-orbit coupling).

*Non-Harmonic Initial Conditions*:

The initial nuclear wavepacket might be quite a bit more diffuse along the I-C-C-I torsion coordinate than can be modeled with the two-conformer harmonic Wigner distribution used here. One possible change in physics that would manifest from a broader torsional wavepacket is lowered recoil between the C involved in the first I dissociation and the second I, which would likely lower the branching ratio for double dissociation.

*Improper FOMO-CASCI Active Space:*

It is possible that the (12e,8o) active space is not accurate across the full part of the potential energy manifold encountered (even discounting spin-orbit coupling effects). We have performed limited XMS-CASPT2 benchmarks to discount this possibility,

*Internal Conversion/Non-Adiabatic Dynamics*:

We have performed very limited AIMS dynamics from initial conditions initiated on $S_1$, and have found that while considerable spawning occurs between TBFs on the nearly-degenerate early $S_1/S_2$ and later $S_0/S_1/S_2$ states (or even higher near-degeneracy for double dissociation), the nuclear part of the wavepacket propagates along essentially the same trajectory. This is expected for an E-type state coalescing to a highly degenerate asymptotic state. Therefore, non-adiabatic dynamics are not expected to provide non-trivial physics for ICs initiated at $S_1$. However, we have not attempted AIMS computations (or adiabatic dynamics) for trajectories initiated from $S_3$ or $S_4$ – the crossings between $S_3/S_4$ and $S_1/S_2$ appear to be quite sharp and electron-transfer like, indicating that most of the population is likely to proceed ballistically along the localized dissociative state, but this would need to be verified by an extensive computational study to draw firm conclusions.

*Rydberg States*:

Rydberg states have not been included in the model. These are believed to lie considerably above the valence states, but may have strong oscillator strength for e.g., two-photon absorption.

*Ionized States*:

Similarly, ionized states have not been considered.

*Differential Dynamical Electron Correlation*:

Differential dynamical electron correlation (possibly in larger basis sets) might move the states relative to one another. Limited XMS-CASPT2 benchmarking described above indicates that this is unlikely within the limited LANL2DZ basis set, but might become an issue in larger basis sets.

### IV. RESULTS

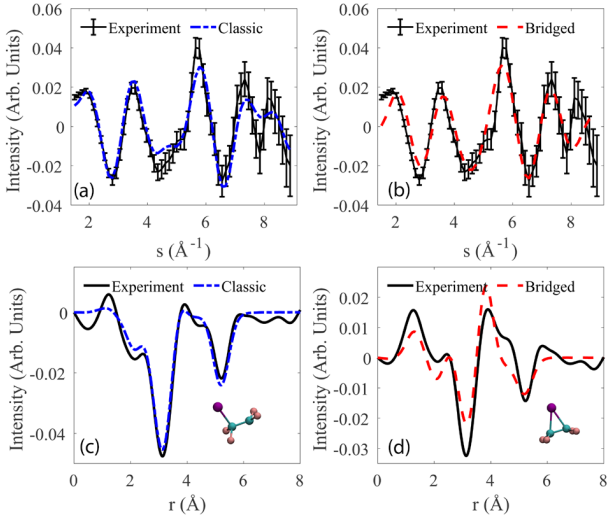

**FIG. 8.** (a,b) Comparison of the experimental $\Delta sM$ at $t=200$ fs (Black) with the classical (Blue-dashed) and bridged (Red-dashed) structures. (c,d) Comparison of the experimental $\Delta fr$ at $t=200$ fs (Black) with the classical (Blue-dashed) and bridged (Red-dashed) structures.

### A. Classical vs Bridged Structure

We first focus on whether the radical takes the classical or bridged structure shortly after dissociation using only the experimental $\Delta sM(s, t=200\ fs)$ which does not require input from theory. We have found that the diffraction data at 200 fs after dissociation matches the classical structure very well and differs significantly from signal expected from the bridged structure.

Figure 8 (a,b) shows the experimental $\Delta sM(s)$ at t = 200 fs along with calculated $\Delta sM(s)$ corresponding to the static classical and bridged structures, respectively. A Chebyshev lowpass filter of order 30 and cutoff frequency 0.001 was applied to the data to reduce noise. The classic structure uses coordinates experimentally measured by Ihee et al. in [1] with UED at a time several picoseconds after the dissociation. The bridged structure [32] was optimized along the lowest lying doublet state at the SA6-CASSCF(7,5) level using the basis sets and corresponding ECP's by Bergner *et al.* [33]. This local minimum was predicted to be visited transiently by the dissociative wave-packet on a time scale of a few tens of femtoseconds, *i.e.* without the formation of stable bridged structures. Figure 8 (a,b) shows that the data matches the classical structure significantly better than the bridged structure, with reduced chi-squared values of 1.8 and 8.9, respectively. While the classical structure fits significantly better than the bridged structure, based on the experimental least squares fit alone we cannot exclude a small percentage of the bridged structure, at a level below 10 percent of the classical structure. Based on the theoretical simulations described earlier in this manuscript, the bridged structure never forms. Figure 8 (c,d) displays the corresponding $\Delta fr(r, t=200\ fs)$. In order to perform the sine transform the experimental values for $s < 1.2$ Å$^{-1}$ were filled in with the respective values for each model (classic and bridged). The structural overlap in real space gives a more intuitive view of the quality of fit for the classic structure over the bridged structure. The relatively small differences between the classic and experimental $\Delta fr$ are thought to arise from the experimental artifacts and from the fact that the molecule is undergoing vibrational motion, which is not included in our model.

### B. Dissociation Dynamics

We now focus on the experimental data to elucidate the dynamics of the dissociation. The $\Delta sM(s < 1.2$ Å$^{-1}$,$t)$ is filled in by a best fit to a set of calculated structures in which only the iodine atom is moved with respect to the rest of the molecule. Diffraction patterns for different structures are generated by increasing one of the C-I distances in steps of 0.01 Å and convolving the resulting $\Delta sM(s,d)$ with a Gaussian to account for temporal-blurring and wavepacket broadening. Each experimental time step is compared with all the calculated structures and the best fit is used to fill in the low s region. This method is applied only for time delays between -200 fs and 50 fs, and reliably captures the increasing C-I distance over this time interval (see Appendix for more details). For all time steps > 50 fs, we use the ground state structure with one iodine removed to fill in the missing data region at small angle.

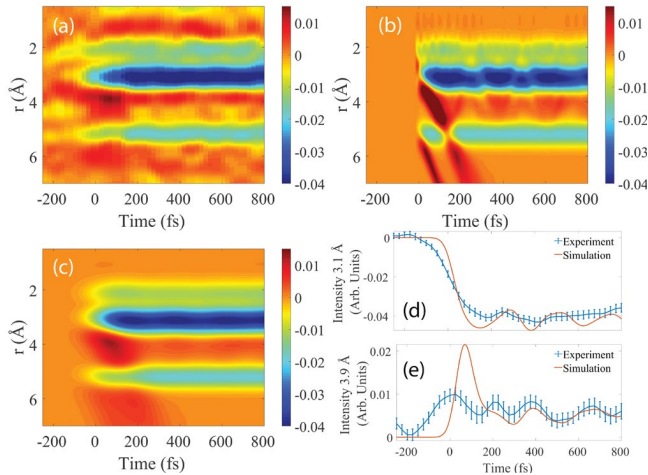

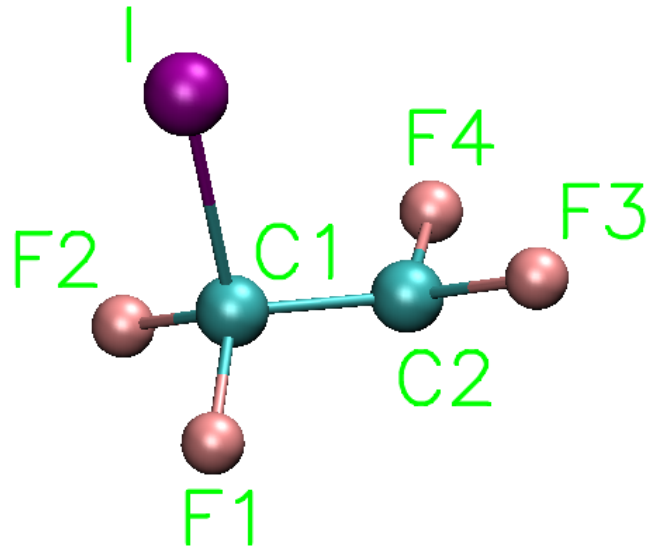

**FIG. 9.** Temporal evolution of $\Delta fr$. (a) Experimental results convolved with a 75 fs FWHM Gaussian filter in time. (b) Simulated results with no convolution (c) Simulated results convolved with 145 fs Gaussian in time to match the temporal resolution of the experiment. (d-e) Lineouts from part (a) and (b), convolved with 75 fs Gaussian to reduce noise. The bleaching signal at 3.1 Å and the dissociating wavepacket at 3.9 Å are visible in the lineouts at t=0. Error bars generated via a bootstrapping method (see Supplementary Material).

Figure 9a shows the experimental $\Delta fr$ as a function of time, displaying the changes in interatomic distances. New distances that form appear as regions with positive counts, while missing distances appear as negative peaks, or bleaches. Starting at t=0, long-lived bleaching signals appear corresponding to the missing C-I, F-I, and I-I distances due to the removal of an iodine atom. There is overall good agreement between experiment (Fig. 9a) and theory (Fig. 9b). The experiment also captures the departing iodine nuclear wavepacket, which can be seen as positive counts in two regions: between 3.5 Å and 4.0 Å and between 5.5 Å and 6.0 Å, at times between zero and 200 fs. The first region corresponds to an increase in distances between the departing I and F1, F2, F3, F4, and C1 (see Fig. 10) which originate at 3.1 Å. The second has two main contributions, the increasing distances that originate at 3.1 Å and a new positive streak that originates at the original I-I distance of 5.1 Å. Figure 9c shows the theory convolved with the experimental time resolution of 145 fs, which shows closer agreement with the experiment. The dissociating wavepacket is no longer visible at distances beyond 6.0 Å, as the wavepacket becomes very broad and the amplitude of the signal falls below the detection level of the experiment.

**FIG. 10.** Transient $C_2F_4I$ with atomic labels.

### C. Coherent motion in the C2F4I radical

In this section we focus on the dynamics that take place in the radical after dissociation. Figure 9d shows a lineout of the $\Delta fr$ signal at the 3.1 Å position (extracted from the data in Fig. 9a and 9b) which corresponds to ground state distances between either iodine atoms and F1, F2, F3, and F4 as well as the long C-I distance (the remaining iodine, I, and C2 or the dissociated iodine and C1). The signal drop is faster in the theory due to the finite time resolution of the experiment. Figure 9e shows a lineout of the $\Delta fr$ at 3.9 Å which is isolated from all of the ground state distances. At $t = 50$ fs, the large positive signal corresponds to the dissociating iodine wavepacket as it passes through this position. In both cases, there are modulations in the amplitude of the $\Delta fr$ that correspond to coherent oscillations in the $C_2F_4I$ radical after the dissociation.

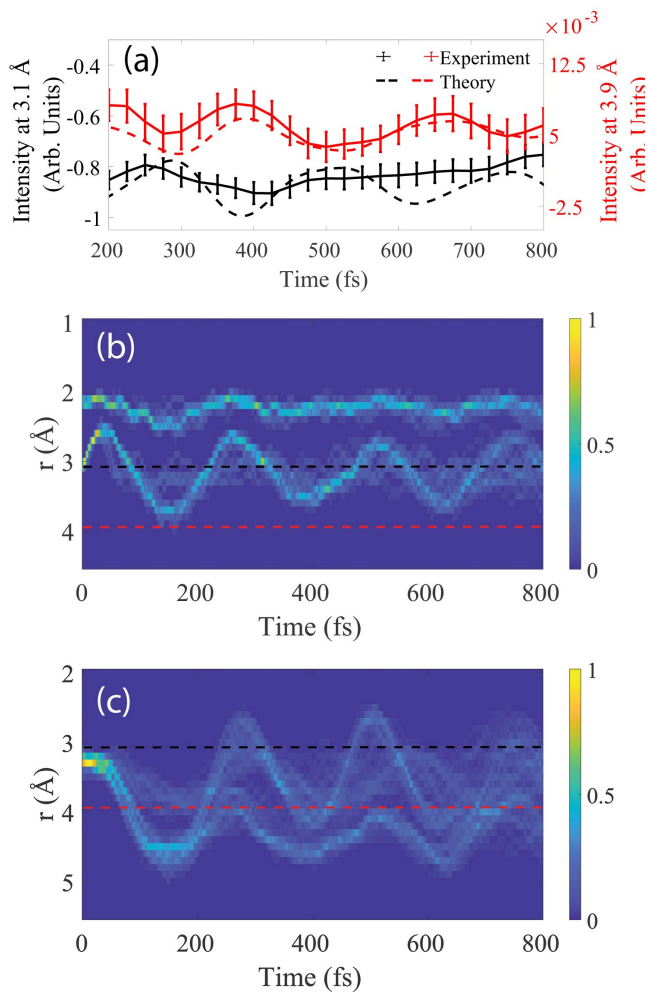

FIG. 11. (a) Lineouts at interatomic distances of 3.1 Å and 3.9 Å show coherent oscillations. (b) Calculated density map of the remaining C-I atomic distances after dissociation. (c) Calculated density map of the remaining long F-I atomic distances after dissociation. The dashed lines at 3.9 Å (Red) and 3.1 Å (Black) in (b) and (c) show the position corresponding to the lineouts in part (a). Error bars were generated via a bootstrapping method.

Figure 11a shows a closeup of the coherent dynamics that take place after 200 fs at the two distances shown in Fig. 3d and Fig. 3e. At the longer distance, where there is less background from competing ground state distances, the amplitude and temporal structure of the measured signal matches very closely with theory. For the shorter distance there is qualitative agreement with theory, however the amplitude of the oscillations is smaller in the experiment. The out of phase oscillation at the two distances is suggestive of a coherent large amplitude vibration as the wavepacket moves from one position to the other.

We now turn to the theory to interpret these signals in terms of nuclear motions. Figure 11b shows calculated density plots of the two C-I distances and Fig. 11c shows the corresponding density plots for the long F-I distance in the $C_2F_4I$ radical from 46 simulated trajectories. Figure 11b shows that the motion is started by the recoil of the carbon atom at dissociation, which results in fast changes of the long C-I distance over the first 30 fs. In Fig. 11c we see that there is no motion of the fluorine atoms over the first 30 fs, followed by large amplitude oscillations. Although, the 3.9 Å lineout in Fig. 11a is in quantitively good agreement with the theory, the oscillations in the experimental lineout at 3.1 Å was found to damp much more rapidly than its theoretically predicted counterpart. We believe this discrepancy to be caused by the different contrast of the vibrations at different distances due to dephasing, which seems to happen more rapidly in the experiment than in the simulation. As shown in Fig. 11b and Fig. 11c, there are multiple distances that contribute to the signal at 3.1 Å, while at the longer distance only the I-F3, I-F4, and I-C2 distances contribute. Based on the theoretical results, we have determined that the oscillations in the data are due to motion of the isolated $CF_2$ group. We consider an atom X at the center of mass of F3 and F4. The oscillations are due to rotations of F3 and F4 about an axis passing through C2 and X and changes in the C1-C2-X angle as well as changes in the I-C1-C2 angle. Movie1 in the Supplementary Information shows a 3D model of the motion based on a representative trajectory from the simulation.

## V. Summary

In summary, we spatially resolved wavepacket dynamics in the photodissociation of $C_2F_4I_2$ using MeV UED and determined that the transient $C_2F_4I$ radical forms with the classical structure within one vibrational period of the relevant bonds. These findings were corroborated by FOMO-CASCI calculations. We have also observed coherent dynamics in the electronic ground state of the radical, with close agreement between experiment and theory.

## VI. Appendix
### A. Legendre Projection

To verify the validity of using the Legendre projections the expected distribution up to $sM_6$ was extracted from the data. Significant signal is seen in $sM_0$ and $sM_2$, minimal signal is seen in $sM_4$, and all other orders are negligible. This is with single photon excitation, which produces photoexcited molecules in an anisotropic distribution. The signal in $sM_2$ is observed to decay in a few picoseconds, as expected from the rotational motion of the molecule [13]. The symmetry found in orders 0, 2, and 4 allow us to "fold" the total intensity pattern into quarters. To fold the images pixels with mirror symmetry about two axes passing through the center of diffraction, one parallel to the laser polarization and the other perpendicular to the polarization, are averaged leaving an image one quarter the size of the original diffraction image. This eliminates discontinuities in the projection due to the areas of the diffraction pattern that have been ignored due to laser leakage and dark current.

### B. Data Analysis

Excessive background from dark current and laser leakage was removed from all images using a standard deviation divided by the mean approach. 120 background images were examined pixel by pixel. Pixels with 15% deviation from the mean were removed from all images. Images were shifted to ensure they are concentric and cropped so that all remaining pixels for a given time step have the same statistical significance. The images are normalized then treated via an iterative method to further reduce noise. Prior to normalizing their intensity, images are radially averaged and pixels that are more than four standard deviations from the mean at each radius are ignored. The total counts in the region between 1.6 Å$^{-1}$ < $s$ < 6.25 Å$^{-1}$ are used as a normalization constant for each full two-dimensional image. After normalization an iterative noise removal method is applied as follows. All steps utilize a four standard deviation cutoff applied to each time step individually. First the cutoff is applied on a pixel by pixel basis for all images in a given time step. Next, all pixels along a given radius from the center are compared for each image individually. This process is repeated once. A time dependent signal is then found for each day separately. The day is then further divided into subsets whose size is determined by the minimum number of images per time point that produce a reasonable time dependent signal, which was 12 to 15 images. To elucidate changes in the intensity difference-images are created. Images with $t$ < -300 fs are averaged and subtracted from all images in the subset. This also helps to reduce experimental background which can vary between time scans. The images in each subset are sequential in time in the lab frame to determine if there is any drift in the time dependent signal within each day. Subsets with significant drift are discarded. Subsets with a reasonably slow drift are shifted and combined into 25 fs time bins based on the rise in the time dependent signal. Time bins that have below 150 total images are discarded. The pixel-radii iteration is applied once more.

Low scattering angles are not present due to the holes in the detection apparatus. Missing low angle experimental data, $s$ < 1.2 Å$^{-1}$ were replaced with theoretical values. As the iodine dissociates the amplitude of the $sM$ at low angles changes due to both the temporal blurring and changes in the interatomic distances. At short times after dissociation, there is a significant effect on the diffraction pattern from the interference of the departing iodine atom with the radical. At later times this interference is no longer visible and the missing data region at low s can be filled using a simple approximation: the diffraction of the parent molecule with one iodine removed. In order to fill in the missing low angle data at early times, structures are created by taking a static ground state molecule and increasing the distance of one C-I pair in discrete steps of 0.01 Å. These theoretical structures are then used to calculate diffraction patterns up to a C-I distance of 12.1 Å. The simulations are then convolved with a Gaussian function with FWHM of 2 Å to model temporal blurring and wavepacket dispersion. The width of the convolution was determined by trial and error with the best result giving a realistic description of the departing iodine atom. The data at each time step between -175 fs and 50 fs is compared with all the calculated diffraction patterns for all the model structures and the one that produces the best fit is used to fill in the small s region. With this method, no time dependent information is introduced in the data. For times before -200 fs the missing data is filled using diffraction from the ground state structure. For times beyond 50 fs, the dissociating iodine atom no longer significantly impacts the diffraction pattern, therefore the same model structure is used for all experimental time steps greater than 50 fs. This is also done to ensure that this fitting in no way affects the oscillations in the $fr$ discussed in the manuscript. The C-I distance extracted from the fitting shows a monotonic increase as expected for a dissociation reaction. Large angle scattering, $s$ > 12 Å$^{-1}$, is not present in the experimental data due to the finite size of the detector. The $sM$ is multiplied by a Gaussian of the form $e^{-ds^2}$ with a damping factor, $d$, of 0.04 (0.06), experiment (theory), to minimize the effect of the truncation at large scattering angles. Remaining background at $s$ > 5 Å$^{-1}$ is removed by fitting the data to a low order polynomial then subtracting off this contribution before transforming.

### C. Experimental Uncertainty

A standard bootstrap approach is used to determine the uncertainty in the experimental data. The complete set of images at each time step is randomly sampled until the total number of images in the sample set matches the complete set. The data analysis is applied to the sample set to produce a sample $\Delta fr$. This process is repeated until the variance in observables in the $\Delta fr$ converge (i.e. adding more samples does not appreciably affect the uncertainty). 100 random samples provided sufficient convergence for all observables. Observables for the mean of the random samples agree with observables for the data set as a whole, providing verification for the validity of the methodology.

### VII. Metadata:

The raw data files used for the theory section of this work are on the MtzGroup fire cluster in the ~parrish/chem/cifr-paper directory. The README files in these directories explain the codes used and the specific structure of the data files.

### VIII. Sample trajectory:

A movie of a sample calculated trajectory used in the analysis is uploaded with the supplemental material.

**Acknowledgements**

The experimental part of this research was performed at SLAC MeV-UED, which is supported in part by the U.S. Department of Energy (DOE) Office of Basic Energy Sciences (BES) SUF Division Accelerator and Detector R&D program, the LCLS Facility, and SLAC under contracts DE-AC02-05-CH11231 and DE-AC02-76SF00515. K.J.W., J.P.F.N.and M.C. were supported by the DOE Office of Basic Energy Science, Chemical Sciences, Geosciences, and Biosciences Division, AMOS program under award DE-SC0014170. R.M.P. and T.J.M. were supported by the DOE Office of Basic Energy Science AMOS program. MG is funded by a Lichtenberg Professorship of the Volkswagen Foundation. T.J.A.W. was supported by the DOE Office of Basic Energy Science, Chemical Sciences, Geosciences, and Biosciences Division, AMOS program.